\documentstyle[12pt]{article} 
\def\prd#1{{\em Phys.~Rev.}~{\bf D#1}\ }
 
\def\plett#1{{\em Phys.~Lett.}~{\bf #1B}\ } 
\def\np#1{{\em Nucl.~Phys.}~{\bf B#1}\ }
\def\deg{\ifmmode{^{\circ}}\else ${^{\circ}}$\fi}

\def\itm#1{\item[$(#1)$]} 
\def\gsim{\,\raisebox{-0.13cm}{$\stackrel{\textstyle>}{\textstyle\sim}$}\,}
 
\def\bi{\begin{itemize}}
\def\ei{\end{itemize}} 
\def\ed{\end{document}}
\def\be{\begin{equation}} 
\def\ee{\end{equation}}
\def\bea{\begin{eqnarray}}
\def\eea{\end{eqnarray}}
\def\lab#1{\label{eq:#1}} 
\def\vev#1{\left<{#1}\right>} 
\def\lam{\ifmmode{\Lambda}\else $\Lambda$\fi}
\def\cond{\vev{\lambda\lambda}}
\def\tc{T_c}
\def\rc{R_c}
\def\nc{N_c}
\def\cn{{\cal N}}
\def\cntot{\cn_{\rm tot}}
\def\cnh{\cn_{\rm hidden}}
\def\tf{t_f}

\def\mpl{M_{\rm Pl}}
\def\tfrac#1#2{{\textstyle\frac{#1}{#2}}}
\def\half{\frac{1}{2}} 

\def\thalf{\tfrac{1}{2}}

\def\gev{\ \mbox{GeV}}
 
\def\mev{\ \mbox{MeV}} 
\def\pri{^{\, \prime}}

\def\ms{m_{3/2}}
\def\mo{M_0}
\def\sh{\hat\sigma}
\def\ash{(A/\sh)}
\def\aosh{\frac{A}{\sh}}
\def\su#1{$\mbox{SU}(#1)$}

\def\eb{\end{thebibliography}}

\def\eqn#1{(\ref{eq:#1})} 
\begin{document}
\begin{titlepage} 
\begin{flushright}  {\sl NUB-3144/96-Th}\\  
{\sl Sept 1996}\\  
hepph/9609506
\end{flushright} 
\vskip 0.5in 
\begin{center}  
{\Large\bf Thermodynamics of Hidden Sector Gaugino Condensation 
in the Expanding Universe}\\[.5in]   
{Haim Goldberg}\\[.1in]  
{\sl Department of Physics}\\   
{\sl Northeastern University}\\ 
{\sl Boston, MA 02115}  
\end{center} 
\vskip 0.4in
\begin{abstract}This work examines the confining-deconfining phase 
transition in a hidden Yang 
Mills sector with scale $\lam \sim 10^{14}$ GeV appropriate to dilaton
stabilization and SUSY-breaking via formation of a gaugino consensate. If the
transition is assumed to take place through homogenous nucleation, then under
reasonable assumptions it is found that a critical
bubble, formed at a temperature which provides enough supercooling, is not 
large
enough to accommodate an adequate number
$(\gsim 100)$ of quanta of the confined phase  (`hidden
hadrons') to allow a consistent thermodynamic description. Thus, a first order
transition in the hidden sector may not be possible in the expanding universe. 
\end{abstract} 
\end{titlepage} 
\setcounter{page}{2} 
\section{Introduction}
An unbroken hidden supersymmetric gauge sector, supporting a gaugino 
condensate
$\cond\sim (10^{14}\ \mbox{GeV})^3,$ can through its coupling to
the dilaton (and other moduli) serve the  related dynamical functions of
providing a superpotential for the dilaton and a mass for the gravitino
 \cite{Nil}.
Because of its possibly central role in the understanding of these 
phenomena, the dynamics associated with this hidden gauge sector may merit
additional study. As a zero temperature field theory, against a Minkowski
space-time background, this has been done (without dilaton) in the original
paper of Veneziano and Yankielowicz \cite{Ven} (see also Amati et al
\cite{Ama}). In the effective theory, the ground state emerges as containing a
gaugino condensate
$\cond$ with residual $Z_N$ symmetry. The changes which evolve with the
introduction of supergravity and the dilaton  are discussed in \cite{Fer}.

In the context of the standard Robertson-Walker cosmology, it is of
interest to examine the transition from the  hot unconfined phase in the hidden
sector to the confined phase.\footnote{It is possible that screening, rather
than confinement, characterizes the phase transition
\cite{Gro}. The only property to be used in this paper is the change in the
degrees of freedom from perturbative non-singlet quanta to gauge-singlet
`hadrons' as one passes through the critical temperature.} Is the transition
first or second order?
If first order, is the transition completed in a manner consistent with the
observed smoothness of the universe? Some possible problems associated with a
transition  described by the evolution of the condensate as a field theoretic
order parameter were discussed in \cite{Gol}. However, the results depended
critically on the K\"ahler potential of the effective condensate field, and
this is certainly not known. Thus, a more phenomenological approach is 
indicated.

In this work, I present a preliminary study of the transition to the confining
phase of the unbroken hidden Yang-Mills sector, under a few well-defined
assumptions:
\bi
\itm{1}There exists a hot, unconfined phase of the hidden sector.
\itm{2}The transition to confining phase is first order, proceeding
through the spontaneous nucleation and expansion of critical bubbles 
of confined
phase.
\ei
Neither of these assumptions is necessarily correct. The aim of the present 
study
is to simply examine the consequences of taking them to be true 

The principal conclusion of this study, under the stated
assumptions, is the following: if the lightest composite particles in the
confined phase (`hidden hadrons') are more massive than $\sim 2\ \tc$ (where
$\tc$ is the critical temperature), then the requirement of sufficient
supercooling in order to complete the phase transition implies a critical
bubble size too small to contain enough ({\em i.e.,}
$>100$) hidden hadrons to allow a meaningful thermodynamic description. This
result is largely traceable to the high value of the transition temperature.
The conclusion will be shown to hold  as well in the presence of
(near) massless vector-like matter fields. Thus  a first
order transition in a {\em cosmological} context may be  unlikely in the hidden
sector. 

\section{Review of Homogeneous Nucleation}
In this section, I will briefly sketch the derivation of the relevant formulae,
initially  following Fuller, Mathews, and Alcock \cite{Ful} in their
discussion of the quark-hadron transition, with some modification for ease of
application to the present situation.

The formation of the confined phase proceeds through the spontaneous nucleation
of critical bubbles of radius $\rc,$ at which the free energy difference
between confined and deconfined phases (pressures $P_{\rm conf}(T)$ and
$P_{\rm deconf}(T),$ respectively)
\be
\Delta F= -\frac{4\pi}{3}\rc^3\ (P_{\rm conf}-P_{\rm deconf}) + 4\pi\sigma\rc^2
\lab{delf}
\ee
has a saddle point at 
\be
\rc(T) = \frac{2\sigma}{P_{\rm conf}(T)-P_{\rm deconf}(T)}\ \ .
\lab{rc}
\ee
Here $\sigma$ is the surface energy density at the interface between the 
phases.
The probability of nucleation of a single bubble is then
\be
p(T)=CT^4\ e^{-\Delta F/T} \ \ .
\lab{p1}
\ee

The critical temperature $\tc$ is define through the coexistence condition in
the infinite volume limit
\be
P_{\rm conf}(\tc)=P_{\rm deconf}(\tc)\ \ .
\lab{coex}
\ee
If the amount of supercooling $\eta\equiv (\tc-T)/\tc$ is small, then one may
expand
\be
P_{\rm conf}(T)-P_{\rm deconf}(T)\simeq L\eta\ \ ,
\lab{delp}
\ee
where 
\be
L=\tc\ \frac{\partial}{\partial T}\left.(P_{\rm deconf}-P_{\rm
conf})\right|_{\tc}
\lab{lat}
\ee
is the latent heat
released per unit volume during the phase change. Combining these results, we
have
\be
p(T)\simeq C\tc^4\ e^{-A(T)}\ \ ,
\lab{p2}
\ee
where
\be
A(T)=\frac{16\pi}{3}\ \frac{\sigma^3}{L^2\tc\eta^2}
\lab{a}
\ee
is the 3-dimensional critical bubble action. The rapid increase of $p(T)$ with
increased supercooling is then explicit.

The condition for the complete nucleation of the universe at time $\tf$ is
\be
\int^{\tf}_{t_c}dt\pri\ p(T\pri(t\pri))\ \frac{4\pi}{3}\ v_s^3\
(\tf-t\pri)^3=1\ \ ,
\lab{nuc1}
\ee
where $v_s\simeq 1/\sqrt{3}$ is the speed of the expanding shock wave of the
confined phase into the metastable unconfined phase.

For small supercooling, one may expand about $t\pri=\tf:$ \cite{Ful}
\be
\ln p(T)=\ln p(T_f) + \left.\frac{d\ln p}{dT}\right|_{T_f}\
\left.\frac{dT}{dt}\right|_{\tf}\ (t-\tf)
\lab{lnp}
\ee
with 
\be
\left.\frac{dT}{dt}\right|_{\tf}=-T_f\ H(T_f)\simeq -\tc H(\tc)\ \ .
\lab{dtdt}
\ee
Thus
\be
p(t)=p(\tf)\ e^{-\lambda(\tf-t)}
\lab{pt}
\ee
with $\lambda=2A(T_f) H(\tc)/\eta_f.$ The condition \eqn{nuc1}\ then
becomes
\be
\frac{8\pi v_s^3}{\lambda^4}\ p(T_f) =1\ \ ,
\lab{nuc2}
\ee
or
\be
\frac{8\pi C}{3\sqrt{3}}\ \left(\frac{\tc}{H(\tc)}\frac{\alpha}{2}\right)^4 =
A^6\ e^A
\lab{nuc3}
\ee
where
\be
\alpha\equiv \left(\frac{16\pi}{3}\ \frac{\sigma^3}{L^2\tc}\right)^{\half}\ \ ,
\lab{alpha}
\ee
and $A\equiv A(T_f).$ Note that
\be
\eta_f=\frac{\alpha}{\sqrt{A}}
\lab{etaf}
\ee
and the Hubble constant at $\tc$ is
\be 
H(\tc)=\frac{1}{\mpl}\ \sqrt{\frac{8\pi\rho(\tc)}{3}}\ \ ,
\lab{htc}
\ee
with
\be
\rho(\tc)=\mbox{total energy density at}\ \tc = \cntot\frac{\pi^2}{30}\
\tc^4\ \ .
\lab{rho}
\ee
Here $\cntot$ is an effective statistical weight for the degrees of
freedom at $\tc.$ If the universe happens to be dominated by a vacuum energy
$\rho_0$ at the time of supercooling, then Eq.~\eqn{rho}\ is just a
redefinition of $\rho_0$ in terms of $\cntot$. 

Combining Eqs.~\eqn{alpha}-\eqn{rho}, the condition \eqn{nuc3}
becomes
\be
A^6\ e^A =
\left(\frac{0.45\ C^{1/4}\ \alpha}{\sqrt{\cntot}}\ \ 
\frac{\mpl}{\tc}\right)^4\ \ .
\lab{nuc4}
\ee
\section{Example of Pure Non-Supersymmetric \su{3} Yang Mills}
As a preliminary for the hidden sector, consider the case of purely
gluonic \su{3}$_c$ (no quarks). Then $\cntot=16,\ \tc\simeq 200$ MeV, and a
lattice-based estimate of $\alpha$ is \cite{Iwa}
\be
\alpha\simeq 0.0065 \pm 0.0015\ \ .
\lab{alphx}
\ee
With $C\sim 1,$ Eq.~\eqn{nuc4}\ yields
\be 
A=124\ \ ,
\lab{a3}
\ee
and Eq.~\eqn{etaf}\ gives $\eta_f= 6\times 10^{-4}.$  From \eqn{rc},
\eqn{delp}, \eqn{alpha}, and \eqn{etaf}, the radius of a
critical bubble is
\be
\rc=\frac{2\sigma}{L\eta_f}=\sqrt{\frac{A}{4\pi}}\
\left(\frac{\tc^3}{\sigma}\right)^{1/2}\ \tc^{-1}\ \ .
\lab{rcf}
\ee
{}From the data of \cite{Iwa}, the surface energy density is not quite scale
invariant with present statistics; nevertheless, it will suffice for the
present application to take \cite{Bro} 
\be
\sigma\simeq 0.025\ \tc^3\ \ ,
\lab{sigma}
\ee
which gives
\be
\rc= 35\ \tc^{-1}\ \ .
\lab{rcx}
\ee
I now come to the point of departure of the present work.

With no quarks, the relevant degrees of freedom in the confined phase are
glueballs with masses $M_i \ \gsim\ 1\gev\gg \tc,$ and number densities
\be
n_i(\tc)\simeq \tc^3\ (2S_i+1) \left(\frac{M_i}{2\pi\tc}\right)^{3/2}\
e^{-M_i/\tc}\ \ .
\lab{ni}
\ee
Then the number of glueballs
in the critical bubble is
\be
\nc =\sum n_i(\tc)\ \ \tfrac{4}{3}\pi\rc^3\ \ .
\lab{nc1}
\ee
Phenomenological studies \cite{Shu}  have suggested that a density of
states
$\tau(M)\propto M^3$ (for each isospin and hypercharge) provides a good fit to
the observed hadron spectrum. I adopt this for the glueball spectrum, and
normalize to 1 state in the interval
$(\mo^2-\thalf\mu^2,\mo^2+\thalf\mu^2),$ where
$\mo$ is the mass of the lightest glueball and $\mu^2$ is the inverse of the
Regge slope. This gives for the spectral density
\be   
\tau(M)=(2/\mo^2\mu^2)\ M^3\ \ ,
\lab{tau}
\ee
and for the number density
\bea
n_c&=&\sum n_i(\tc)=2(2\pi)^{-3/2}\mo^{-2}\mu^{-2}\int_{\mo}^{\infty} dM\ M^3\
(M/\tc)^{3/2}\ e^{-M/\tc}\ \ \tc^3 \nonumber\\
&\equiv& 2(2\pi)^{-3/2}\ (\tc/\mu)^2\  f(\mo/\tc)\ \ \tc^3\ \ .
\lab{nc2}
\eea
Combining \eqn{rcx}, (\ref{eq:nc1}), and (\ref{eq:nc2}), one finds
\be
N_c=22,800\ (\tc/\mu)^2\  f(\mo/\tc)\ \ .
\lab{nc3}
\ee
A thermodynamic description, and
hence a first order phase transition, will be {\em cosmologically} viable only
for
$N_c\gg 1,$ say
$N_c>100.$ Once $(\mu/\tc)$ is specified, this will translate via
(\ref{eq:nc3}) to a bound on
$\mo.$ Since $\mu^2\sim 1 \ \gev^2,$ and $\tc\simeq 250\ \mev,$ one finds
\be
\mo\le 8\ \tc\simeq 2000\ \mbox{MeV}\ \ .
\lab{mbound}
\ee

Before continuing, I will comment briefly on the question of 
possible temperature
dependence of the glueball masses near $\tc.$ A recent theoretical calculation
\cite{Sug} in the context of the dual Ginzburg-Landau theory shows 
that there is
some reduction in mass of the gauge singlet QCD monopole at $\tc,$ although the
effect is totally dependent on an assumed (and unknown) temperature dependence
of the dual Higgs quartic coupling. There is no basis for supposing that the
glueball mass goes to zero (or is even much reduced) at $\tc$ in a strongly
first order phase transition, and I will simply reinterpret
\eqn{mbound}  as a bound on an effective glueball mass, with the expectation
that it does not differ significantly from the zero temperature mass. 

I now turn to examine the implication  of these ideas
when applied to the hidden gauge sector which is  relevant to gaugino
condensation and SUSY-breaking.
\section{The Hidden Sector: Pure SUSY Yang-Mills}
There are several significant differences between the hidden sector
SUSY-Yang-Mills theory and the non-SUSY SU(3)$_c$ theory just discussed:
\bi
\item The theory is supersymmetric: there are Majorana fermions in the adjoint
representation, and only rudimentary lattice results are available for such a
theory \cite{Mon}.
\item The vacuum properties are totally unlike those in ordinary QCD \cite{Ama}.
\item The gauge group is either much larger than SU(3), or is manifest at  a
Kac-Moody level $k\ge 2,$ in order that strong 
coupling sets in at a scale $\sim
10^{14}$ GeV.
\ei
A question which immediately arises as a consequence of these differences is
whether the transition is first order. The simplest approach here is to assume
that it is first order as a working hypothesis, and examine the consequences. 

Since the interface energy $\sigma$ and the parameter $\alpha$ are
completely unknown for the hidden sector, the nucleation condition
\eqn{nuc4}\ must be rewritten in a suitable manner. I assume that the
specific entropy of the hadronic phase is much less than that of the gauge
phase, so that I take for the latent heat
\be
L=\tc\ \left.\frac{dP_{\rm deconf}}{dT}\right|_{T=\tc}=4\
\frac{\pi^2}{90}\ \cnh\ \tc^4\ \ .
\lab{lat1}
\ee
It is also convenient to set 
\be
\sh\equiv \sigma/\tc^3\ \ ,
\lab{sigh}
\ee
so that the condition \eqn{nuc4}\ becomes
\be
\left(\aosh\right)^6\ e^A=\left(\frac{4.8\ C^{1/4}}{\cnh\sqrt{\cntot}}\
\frac{\mpl}{\tc}\right)^4\ \ .
\lab{cond}
\ee

What is $\tc?$
For zero cosmological constant, the gravitino mass is given in terms of the
effective superpotential $W$ and Kahler $K$ by
\be
\ms=e^{K/2}|W_{eff}|/M^2\ \ ,
\lab{m32}
\ee
where $M=\mpl/\sqrt{8\pi}.$ In the effective theory, with a simple gauge group
and no matter fields, one obtains after integrating out the gaugino condensate
\cite{deC}
\be
W_{eff}=-\frac{b}{6{\rm e}}\ M_{\rm string}^3\ e^{-3S/2b}\ \ ,
\lab{weff}
\ee
where $b=\beta(g)/g^3=3N/16\pi^2$ for \su{N}, Re $S=1/g_{\rm string}^2\simeq
2.0$ at the correct minimum for the dilaton field $S.$ $M_{\rm string}$ 
sets the
scale for the logarithmic term in the condensate
superpotential. The \su{N} theory becomes strong $(g^2/4\pi=1)$ at a scale
\be
\Lambda=M_{\rm string}\ e^{-S/2b}\ \ ,
\lab{lambda}
\ee
so that 
\be
\ms=e^{K/2}\ \frac{b}{6{\rm e}}\ \frac{\Lambda^3}{M^2}\ \ .
\lab{m32a}
\ee
For $\ms\simeq 10^3\ \gev,$ one obtains
\be 
\Lambda=e^{-K/6}\ N^{-1/3}\cdot 1.7\times 10^{14}\ \gev\ .
\lab{lambda1}
\ee
As a heuristic example, I will choose as the hidden gauge group SU(6), which is
consistent with Eqs.~\eqn{lambda}\ and \eqn{lambda1}\ for 
$M_{\rm string}\simeq 10^{18}\ \gev, \ e^{-K/6}\simeq 1.$ With $\tc\simeq
\Lambda\simeq 10^{14}\ \gev,\
\cnh=2\left(\frac{15}{8}\right)\left(6^2-1\right)=131.25,$\ $\cntot=
\cn(Standard
\ Model)+\cnh=213.75+131.25=345,$ and $C^{1/4}\simeq 1,$ the
constraint Eq.~\eqn{cond} becomes
\be
A + 6\ln\ash = 21\ \ .
\lab{cond1}
\ee
For consistency, we must require that $A$ not be small, so that for $A\ge 1,$
one obtains an upper bound 
\be
A/\sh\le 28\ \ .
\lab{ash}
\ee
As before, I now proceed to calculate the number of (hidden) hadrons in a
critical bubble. 

With the same spectrum of glueballs as for the non-SUSY example of the last
section (Eq.~\eqn{nc2} with a factor of 4 included for the supermultiplet),
and Eq.~\eqn{rcf} for the bubble radius, one finds
\be
N_c=(\sqrt{6}/\pi^2)\ 
(\tc/\mu)^2\ f(\mo/\tc)\ \ash^{3/2}
\lab{nch}
\ee 
where again $1/\mu^2$ is the Regge slope for the hidden sector glueballs.
As previously, the requirement that the cosmological description be
thermodynamically viable requires that $N_c\gg 1,$ which I take to mean
$N_c\gsim 100.$ With the use of \eqn{ash} this devolves to a constraint on
$f(\mo/\tc),$ and hence on
$\mo:$ for $\mu/\tc\ge 3,$ I find
\be 
\mo\le 1.4\ \tc\ \ ,
\lab{mbound1}
\ee
which is of dubious credibility since we expect $\mo>\mu.$ For $\mu/\tc\ge
2,$ the bound is
\be 
\mo\le 2.0\ \tc\ \ ,
\lab{mbound2}
\ee
which is marginally possible. Thus, the conclusion at this point is that a
bubble description for the first order transition is barely possibly (for
$\mo\simeq \mu\simeq 2\ \tc),$ but seemingly unlikely.
\section{Hidden Sector with Matter Fields}
Suppose that the hidden sector contains $N_f< N$ flavors of 
vector-like pairs of
chiral superfields $Q+\overline{Q}.$ If these are massive ($M_Q\gg \Lambda_N,$
the
\su{N} confining scale), then they effectively decouple from the dynamics
discussed in this paper. If they are massless (or nearly massless), then the
discussion in Refs.~\cite{Ama} and \cite{Aff} is germane: the existence of flat
directions $v_{ir}=v_r\delta_{ir} (i=\mbox{color}, r=\mbox{flavor})$ in  the
field space of the $Q+\overline{Q}$ breaks the symmetry to \su{N\pri},
$N\pri\equiv N-N_f,$  at the scale
$v.$ Between
$v$ and $\Lambda_{N\pri},$ the effective massless degrees of freedom are the
Goldstone bosons of the broken flavor symmetry and the \su{N\pri} gauge
degrees of freedom. (I assume that it is $\Lambda_{N\pri}$ which establishes
the condensate scale of interest in this work.) If the 
Goldstones are in thermal
equilibrium with the \su{N\pri} fields, then they would contribute to the
pressure of the critical bubble with high number density, and the problems
encountered earlier would be alleviated. I will now show that the 
Goldstones are
{\em not} in thermal equilibrium with the \su{N\pri} fields, and thus 
the bubble
is transparent to their existence. 

Thermal equilibrium requires that the  ratio $\Gamma/H$ be $>1$ during the era
of interest, where $\Gamma$ is the reaction rate of the Goldstones in the
\su{N\pri} plasma. The coupling of a Goldstone to a  pair of \su{N\pri}
gluons is  $\frac{g^2}{32\pi^2v\sqrt{N\pri}},$ and the cross section in the
plasma
is easily calculated:
\be
\sigma\sim \frac{g^2}{4\pi}\ \left(\frac{g^2}{32\pi^2v}\right)^2\ \ ,
\lab{cross}
\ee
independent of temperature and $N\pri.$ The Hubble constant $H\simeq
\sqrt{\cntot}\ T^2/\mpl,$ so that
\be
\frac{\Gamma}{H}=\frac{\sigma n v_G}{H}\simeq
10^{-8}\ \frac{\cn_{N\pri}}{\sqrt{\cntot}}\ \frac{T\mpl}{v^2}\ \ ,
\lab{gamh1}
\ee
where $n=$ plasma number density, $v_G=$ Goldstone velocity. For the \su{6}
example $(N\pri=6),$ with $T=\tc\simeq 10^{14}, \cn_{N\pri}=131.25,
\cntot=345,$ one finds
\be
\Gamma/H \simeq 0.007\ (v/\tc)^{-2}\ll 1\ \ 
\lab{gamh2}
\ee
for any $v\ge \tc.$ Thus, the Goldstones decouple from the \su{N\pri} plasma,
and do not contribute to the bubble dynamics. 
\section{Summary and Conclusions}
\bi
\itm{a}The  field theoretic description of hidden gluino
condensates must imply a parallel thermal/cosmological description of the phase
transition beween  the unconfined and confined phases of the 
unbroken Yang-Mills
theory. This work has examined the conditions under which a
first order transition in terms of classical bubble nucleation is possible. The
result found is that  only if the mass of the confined phase
glueballs (and superpartners) is very near to $ 2 \tc$ , are critical
bubbles  large enough to contain an adequate number of quanta of the confined
phase particles to  satisfy the thermodynamic conditions for a first order
transition in the expanding universe.  This is true whether or not
there are Goldstones associated with massless vector-like matter fields.
The highly restrictive conditions on glueball mass leads one to  
question whether a first order
transition is possible in the expanding universe. 
\itm{b}If a first order transition is not feasible, then a field theoretic
description of a second (or higher) order transition may be of interest. This
entails some difficulty with the Witten index theorem\cite{Wit}: at the
critical temperature, the order parameter (presumably the gaugino condensate)
must change in a continuous manner from  zero to a non-zero value. However, the
index for the final state is $N$ (for \su{N}),  whereas for 
the initial state it
is (presumably) zero. Resolution of this problem will no doubt involve some
non-trivial input to the effective  theory.
\itm{c}The critical input to the present analysis is the ratio $\mo/\tc,$
where $\mo$ is the mass of the lightest glueball. It would be extremely 
useful to
have some indication, possibly from a lattice study,  of this quantity. A
continued pursuit of SUSY on the lattice, following the initial effort in
Ref.~\cite{Mon} would be very welcome.\ei
\clearpage
\noindent{\bf Acknowledgement}

This work was supported in part by Grant No. PHY-9411546 from the National
Science Foundation.

\ed